\documentclass[manuscript,screen]{acmart}

\AtBeginDocument{%
  }

\setcopyright{acmcopyright}
\copyrightyear{2023}
\acmYear{2023}
\acmDOI{XXXXXXX.XXXXXXX}

\acmPrice{15.00}
\acmISBN{978-1-4503-XXXX-X/18/06}

\usepackage{bm}
\usepackage{subcaption}

\usepackage{multirow}
\usepackage{mathtools,amssymb,euscript,yfonts,latexsym}
\usepackage[capitalise]{cleveref}
\usepackage{makecell}

\usepackage{color,colortbl}
\usepackage{textcomp}

\definecolor{LightCyan}{rgb}{0.8,1,1}
\definecolor{Gray}{gray}{0.9}

\renewcommand{\paragraph}[1]{\vspace{1.25mm}\noindent\textbf{#1}}

\makeatletter
\usepackage{xspace}
\def\@onedot{\ifx\@let@token.\else.\null\fi\xspace}
\DeclareRobustCommand\onedot{\futurelet\@let@token\@onedot}

\begin{document}


\title{A Survey on Audio Diffusion Models: Text To Speech Synthesis and Enhancement in Generative AI}


\author{Chenshuang Zhang}
\affiliation{%
  \institution{KAIST}
  \country{South Korea}
}
\email{zcs15@kaist.ac.kr}

\author{Chaoning Zhang}
\authornote{Correspondence author.}
\affiliation{%
  \institution{Kyung Hee University}
  \country{South Korea}
}
\email{chaoningzhang1990@gmail.com}

\author{Sheng Zheng}
\affiliation{%
  \institution{Beijing Institute of Technology}
  \country{China}
}
\email{zszhx2021@gmail.com}

\author{Mengchun Zhang}
\affiliation{%
  \institution{KAIST}
  \country{South Korea}
}
\email{zhangmengchun527@gmail.com}

\author{Maryam Qamar}
\affiliation{%
  \institution{Kyung Hee University}
  \country{South Korea}
}
\email{maryamqamar@khu.ac.kr}

\author{Sung-Ho Bae}
\affiliation{%
  \institution{Kyung Hee University}
  \country{South Korea}
}
\email{shbae@khu.ac.kr}

\author{In So Kweon}
\affiliation{%
  \institution{KAIST}
  \country{South Korea}
}
\email{iskweon77@kaist.ac.kr}

\renewcommand{\shortauthors}{Zhang et al.}





\begin{abstract}
Generative AI has demonstrated impressive performance in various fields, among which speech synthesis is an interesting direction. With the diffusion model as the most popular generative model, numerous works have attempted two active tasks: text to speech and speech enhancement. This work conducts a survey on audio diffusion model, which is complementary to existing surveys that either lack the recent progress of diffusion-based speech synthesis or highlight an overall picture of applying diffusion model in multiple fields. Specifically, this work first briefly introduces the background of audio and diffusion model. As for the text-to-speech task, we divide the methods into three categories based on the stage where diffusion model is adopted: acoustic model, vocoder and end-to-end framework. Moreover, we categorize various speech enhancement tasks by either certain signals are removed or added into the input speech. Comparisons of experimental results and discussions are also covered in this survey.
\end{abstract}

\keywords{Survey, Generative AI, AIGC, Diffusion model, Text to speech, Speech enhancement, Speech synthesis}

\maketitle

\section{introduction}

Recently, generative AI has attracted unprecedented attention in both academia and industry area~\cite{zhang2023complete}. Aiming at generating content in different modalities~\cite{zhang2023complete}, typical tasks in  generative AI include but are not limited to chatbot (e.g., ChatGPT~\cite{zhang2023ChatGPT}), text-to-image synthesis~\cite{zhang2023text} and text-to-speech synthesis~\cite{jeong2021diff,zhang2023complete}. It should be noted that the success of the diffusion model in computer vision has inspired numerous works for speech generation. This work conducts a survey on audio diffusion model for the recent progress in speech synthesis with the focus on text-to-speech synthesis and speech enhancement.

Speech enables humans to express their thoughts and communicate with each other accurately and efficiently. Therefore, speech synthesis is an indispensable component in modern AI system. Specifically, the text-to-speech and speech enhancement task are two main active tasks, which generates a speech from a given text and enhances the quality of an existing speech, respectively. The development of text-to-speech task can be roughly divided into three stages~\cite{tan2021survey}: early works (e.g., formant synthesis~\cite{seeviour1976automatic,klatt1980software,klatt1987review} and concatenative  synthesis~\cite{olive1977rule,moulines1990pitch,hunt1996unit}), statistical parametric speech synthesis (SPSS)-based methods~\cite{yoshimura1999simultaneous,tokuda2000speech,tokuda2013speech}, and neural network-based stage~\cite{ze2013statistical,zen2015unidirectional,wang2017tacotron}. More recently, diffusion model has attracted great attention in multiple fields (e.g., computer vision) and has also been leveraged into the text-to-speech task~\cite{jeong2021diff,huang2022prodiff,kang2022any,yang2022norespeech}. Speech enhancement ~\cite{nuthakki2022literature} is another active research field in speech area, which generates speech with a speech signal as input. Common speech enhancement tasks include speech denoising~\cite{lu2021study}, dereverberation~\cite{welker2022speech,richter2022speech}, and speech super-resolution~\cite{kuleshov2017audio,lim2018time}.

\textbf{Related survey works.} Multiple works have conducted a survey on diffusion models covering all fields~\cite{yang2022diffusion,cao2022survey}. Recently, there are also some field-specific survey works, including text-to-image diffusion models~\cite{zhang2023text}, graph diffusion models~\cite{zhang2023graph_survey}. Complementary to the above works, this work conducts a survey on audio diffusion models. From the perspective of AI-generated content (AIGC), this survey is also related to generative AI (see~\cite{zhang2023complete} for a survey) and ChatGPT (see~\cite{zhang2023ChatGPT} for a survey). From the perspective of speech synthesis, our work is related to multiple survey works\cite{tabet2011speech,ning2019review,tan2021survey,nuthakki2022literature}. While ~\cite{tabet2011speech} reviews text-to-speech works, it mainly highlights the digital signal processing components. Other works~\cite{tan2021survey,ning2019review,nuthakki2022literature} review speech synthesis based on deep learning, which give us an understanding of the development of neural text-to-speech or speech enhancement. However, as far as we know, our survey is the first work that focuses on reviewing the recent progress in diffusion-based speech synthesis. Specifically, we first briefly introduce the background of audio and speech in section~\ref{sec:background}. Section~\ref{sec:text_to_speech} and section~\ref{sec:enhancement} discusses recent diffusion-based progresses on text-to-speech synthesis and speech enhancement, respectively. Section~\ref{sec:conclusion} further summarizes the entire paper. 

\section{Background}
\label{sec:background}

\subsection{Background on audio and speech}

 An audio wave in the air is the vibration of its molecules, through which the sound travels. An audio waveform represents the vibration displacement over time, and its strength is indicated by the amplitude~\cite{tzanetakis2000marsyas}. In essence, an audio waveform is a mixture of frequencies~\cite{vasquez2019melnet,dobrynin2010time}, and therefore, audio analysis often starts with transforming the audio from its raw waveform in the time domain to the spectrogram in the time-frequency domain. This is achieved by segmenting the audio into windows, for which a short-time Fourier Transform (STFT)~\cite{magron2018model} calculate its magnitude for each frequency. The STFT is repeated for every window along the time direction, resulting in a map of complex values with two dimensions: X-axis representing frame (time) and Y-axis representing frequency. The complex values on the map can be further converted into their absolute values (or amgnitude) and phase. Since human are not sensitive to the frequency equally, a mel-scale is often adopted to transform the spectrogram (magnitude map) into Mel-spectrogram~\cite{liu2009temporal,molau2001computing}. Mel-spectrogram is widely applied in numerous speech-related tasks~\cite{jeong2021diff,popov2021grad,liu2022diffgan,yang2022norespeech}, including text-to-speech and speech enhancement discussed in this work.

\subsection{Background on diffusion model}

Before introducing the application of diffusion model in speech synthesis, we first briefly revisit how diffusion model works. As one of the generative model, diffusion model can be traced back to Deep Diffusion Model (DPM)~\cite{sohl2015deep} inspired by non-equilibrium thermodynamics. The process of this model is gradually adding noise to a prior distribution and then reversing it to get synthetic data. Denoising diffusion Probabilistic models (DDPM)~\cite{ho2020denoising} improved DPM~\cite{sohl2015deep} and obtained high-resolution images, which attracted a lot of attention. There is another branch of diffusion models similar to DDPM, the score-based generative model~\cite{song2019generative,song2020improved}. Although formulated from different perspectives, DDPM~\cite{ho2020denoising} and score-based generative model~\cite{song2019generative,song2020improved} turn out to be equivalent in certain settings. Before we introduce how diffusion model is applied in speech synthesis field, we take DDPM as an example and revisit how DDPM works as follows.

DDPM~\cite{ho2020denoising} consists of a forward process and a reverse process. During the forward process, DDPM adds noise to a clean image step by step until the image is destroyed.  Define the data distribution 
$\mathbf{x}_0 \sim q(\mathbf{x}_0)$ and the latent variable $\mathbf{x}_t$ at step $t$, the approximate posterior is fixed to a Markov chain and defined as follows (quoted from ~\cite{ho2020denoising}):

\begin{align}
q(x_{1:T} | x_0) := \prod_{t=1}^T q( x_t | x_{t-1} ), \label{eq:forwardprocess_1}
\end{align}

\begin{align}
q(x_t|x_{t-1}) := \mathcal{N}(x_t;\sqrt{1-\beta_t} x_{t-1},\beta_t I) \label{eq:forwardprocess_2}
\end{align}

In the reverse process, DDPM aims to recover the $\mathbf{x}_0$ from destroyed $\mathbf{x}_T$. This is trained by estimating the noise during training.

\section{Text-to-speech synthesis}
\label{sec:text_to_speech}

Text-to-speech, also known as speech synthesis, generates speech from text, which has been been used in numerous applications. 

\subsection{Overview of the text-to-speech development}

\begin{figure}[!htbp]\centering
\includegraphics[width=0.6\linewidth]{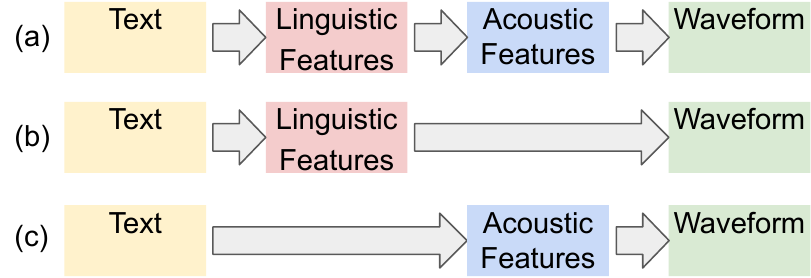}\\
\caption{Development of text-to-speech frameworks. (a) Three-stage framework (b) One branch of two-stage framework that generates waveform from linguistic features directly (c) Another branch of two-stage framework that generates acoustic features from text directly.}
\label{fig:frameworks}
\end{figure}

\begin{table*}[!htbp] \centering
\caption{Recent progress of text-to-speech diffusion models}
\label{tab:diffusion_papers}
\resizebox{0.8\textwidth}{!}{
\begin{tabular}{l|l|l|l}
\hline
   Stage  & Category  &   Methods         & Resources  \\
\hline
Acoustic model & Pioneering work & Diff-TTS~\cite{jeong2021diff}        &   \\
& & Grad-TTS~\cite{popov2021grad}               & \href{https://github.com/huawei-noah/Speech-Backbones/tree/main/Grad-TTS}{Code} / \href{https://grad-tts.github.io/}{Project}  \\
\cline{2-4} 
& Efficient acoustic model &  ProDiff~\cite{huang2022prodiff}    & \href{https://prodiff.github.io/}{Project}   \\
 & & DiffGAN-TTS~\cite{liu2022diffgan}       & \href{https://github.com/keonlee9420/DiffGAN-TTS}{Code}   \\
\cline{2-4} 
& Adaptive multi-speaker model  & Grad-TTS with ILVR~\cite{levkovitch2022zero}          & \href{https://alonlevko.github.io/ilvr-tts-diff}{Project}   \\
& &  Grad-StyleSpeech~\cite{kang2022any}    & \href{https://nardien.github.io/grad-stylespeech-demo/}{Project}   \\
& &  Guided-TTS~\cite{kim2021guided}    &  \\
& &  Guided-TTS 2~\cite{kim2022guided}      & \href{https://ksw0306.github.io/guided-tts2-demo/}{Project} \\
\cline{2-4} 
&With discrete latent space & Diffsound~\cite{yang2022diffsound}      & \href{http://dongchaoyang.top/text-to-sound-synthesis-demo/}{Project}   \\
 & & NoreSpeech~\cite{yang2022norespeech}              &  \\

\cline{2-4} 
& Fine-grained control 
& EmoDiff~\cite{guo2022emodiff}             & \href{https://cantabile-kwok.github.io/EmoDiff-intensity-ctrl/}{Project}    \\
\hline 
Vocoder & Pioneering work& WaveGrad~\cite{chen2020wavegrad}  & \href{https://github.com/ivanvovk/WaveGrad}{Code} / \href{https://wavegrad.github.io/}{Project}   \\
& &  DiffWave~\cite{kong2020diffwave}    & \href{https://diffwave-demo.github.io/}{Code}  \\
\cline{2-4}

& Efficient vocoder & BDDM~\cite{lam2022bddm}   & \href{https://github.com/tencent-ailab/bddm}{Code}   \\
&  & InferGrad~\cite{chen2022infergrad}       & \\
& &  WaveFit~\cite{koizumi2022wavefit}                & \href{https://google.github.io/df-conformer/wavefit/}{Project}   \\
\cline{2-4}
& Statistical improvement &DDGM~\cite{nachmani2021denoising}    &  \\
 & & PriorGrad~\cite{lee2021priorgrad} &  \href{https://speechresearch.github.io/priorgrad/}{Project}  \\
& & It{\^o}Wave~\cite{wu2022itowave}  & \href{https://wushoule.github.io/ItoAudio/}{Project}  \\
& & SpecGrad~\cite{koizumi2022specgrad}    &  \\
\hline
End-to-end&Pioneering work &  WaveGrad 2~\cite{chen2021wavegrad}       &\href{https://github.com/keonlee9420/WaveGrad2}{Code} / \href{https://mindslab-ai.github.io/wavegrad2/}{Project}  \\
&&  CRASH~\cite{rouard2021crash}  & \href{https://crash-diffusion.github.io/crash/}{Project}   \\
\cline{2-4}
&Efficient model& FastDiff~\cite{huang2022fastdiff}  & \href{https://github.com/Rongjiehuang/FastDiff}{Code} / \href{https://fastdiff.github.io/}{Project}   \\
\cline{2-4}
& Further improvements  & DAG~\cite{pascual2022full}  \\
& &It{\^o}n~\cite{shi2022iton} & \href{https://iton007.github.io/}{Project}   \\
\hline
\end{tabular}}
\end{table*}

\textbf{From three-stage to two-stage framework.} The development of text-to-speech has undergone through a shift from a three-stage framework to a two-stage framework, as shown in Figure~\ref{fig:frameworks}. Before applying neural networks, statistical parametric speech synthesis (SPSS) was a popular method~\cite{yoshimura1999simultaneous,yoshimura2002simultaneous,tokuda2000speech,zen2009statistical,tokuda2013speech} consisting of three stages. As shown in Figure~\ref{fig:frameworks} (a), the text input is first converted to linguistic features,  then acoustic features, and to the waveform in the last stage. Common acoustic features include mel-cepstral coefficients~\cite{fukada1992adaptive}, Mel-generalized coefficients~\cite{tokuda1994mel}, F0~\cite{kawahara1999restructuring} and band aperiodicity~\cite{kawahara2001aperiodicity}.   Neural networks have brought a paradigm shift from three stages (Figure~\ref{fig:frameworks} (a)) to two stages (Figure~\ref{fig:frameworks} (b) and Figure~\ref{fig:frameworks} (c)). One branch of two-stage framework makes a deep vocoder directly generates waveform from linguistic features (Figure~\ref{fig:frameworks} (b)), 
such as WaveNet~\cite{oord2016wavenet}, Parallel Wavenet~\cite{oord2018parallel}, DeepVoice 1~\cite{arik2017deep}  DeepVoice 2~\cite{gibiansky2017deep}, HiFi-GAN~\cite{binkowski2019high}. Currently, another two-stage framework is more dominant as shown in Figure~\ref{fig:frameworks} (c), which directly generates Mel-Spectrogram form of acoustic features from text with a single deep acoustic model, including DeepVoice 3~\cite{ping2017deep}, TransformerTTS~\cite{li2019neural}, Fast Speech 1~\cite{ren2019fastspeech} and Speech 2~\cite{ren2020fastspeech}.

\textbf{Overview of diffusion-based methods.} Most recent text-to-speech works on diffusion model follow the two-stage framework in Figure~\ref{fig:frameworks} (c), which first generate acoustic features with a acoustic models, and then output waveform with a vocoder. Another branch of work  attempts to solve the text-to-speech task in an end-to-end manner. We summarize recent work in Table~\ref{tab:diffusion_papers}, and will introduce  each stage within the framework, respectively.

\subsection{Acoustic model}
Acoustic model that transforms a text to acoustic features is a core component in the the task of text to speech. A summary of representative work that apply diffusion model to acoustic model is shown in Table~\ref{tab:acoustic_results}. 

\begin{table*}[!htbp] \centering 
\caption{Experimental results of acoustic model based on diffusion model}
\label{tab:acoustic_results}
\resizebox{0.7\textwidth}{!}{
\begin{tabular}{l|c|l|c|l|c|cccccccccc}
\hline
Methods      & Stage          &    Dataset   &  MOS (↑)& RTF (↓) &SMOS (↑) & CER (↓)\\
\hline
Diff-TTS~\cite{jeong2021diff}     & Acoustic model & LJSpeech &   4.337   & 0.035& - &  -   \\
Grad-TTS~\cite{popov2021grad}     & Acoustic model &   LJSpeech   & 4.44 & 0.012 & - &  -     \\
ProDiff~\cite{huang2022prodiff}  & Acoustic model & LJSpeech & 4.08& 0.04 & - & -  \\
\hline

NoreSpeech~\cite{yang2022norespeech}   & Acoustic model & LibriTTS&   4.11  & - &4.14&- \\
Grad-TTS with ILVR~\cite{levkovitch2022zero}         & Acoustic model & LibriTTS&3.96 & -&-&-  \\
Grad-StyleSpeech~\cite{kang2022any}  & Acoustic model & LibriTTS &4.18 & - &3.83  &2.79    \\
Guided-TTS 2~\cite{kim2022guided} & Acoustic model & LibriTTS &   4.25 & - &3.51  &0.8 \\
\hline

Grad-StyleSpeech~\cite{kang2022any}  & Acoustic model & VCTK & 4.13  & - & 3.95&2.49 \\
Guided-TTS 2~\cite{kim2022guided} & Acoustic model & VCTK &      4.23       & - & 3.39&0.81  \\
\hline

\end{tabular}}
\end{table*}

\subsubsection{Pioneering works}
\ 

In speech synthesis systems, an acoustic model converts the text into acoustic features (e.g., Mel-spectrogram). Diff-TTS~\cite{jeong2021diff} is the first work that applies DDPM to  el-spectrogram generation. In Diff-TTS~\cite{jeong2021diff}, a text encoder first extracts the contextual information, which is then aligned with the length predictor and duration predictor. After that, a decoder with the similar architecture as DiffWave~\cite{kong2020diffwave} is trained to generate mel-spectrogram with DDPM. Specifically, DDIM is leveraged for accelerated sampling. Different from Diff-TTS~\cite{jeong2021diff} with DDPM, Grad-TTS~\cite{popov2021grad} is formulated on stochastic differential equation (SDE)~\cite{song2019generative} and adopts U-net from WaveGrad~\cite{chen2020wavegrad} as the mel-spectrogram generation decoder.  Grad-TTS~\cite{popov2021grad} also provides the possibility of an end-to-end text-to-speech pipeline by replacing  mel-spectrogram with waveforms as the decoder output.

\subsubsection{Towards efficient acoustic model}
\ 

\textbf{Acceleration with knowledge distillation.} According to ProDiff~\cite{huang2022prodiff}, estimating the gradient of data density (gradient-based parameterization) is dominant in previous diffusion models, which requires hundreds or  thousands of iterations for high quality generation ~\cite{huang2022prodiff}. In order to reduce the iterations as well as maintain quality, ProDiff~\cite{huang2022prodiff} adopts the \textit{generator-based parameterization} that directly estimates the clean data. Moreover, ProDiff~\cite{huang2022prodiff} proposes to reduce the data variance by knowledge distillation that optimizes the student model to mimic an N-step  DDIM teacher model with N/2 steps. Experimental results show that ProDiff~\cite{huang2022prodiff} is the first time to make the diffusion models  applicable to interactive, real-world speech synthesis applications at a low computation.

\textbf{Acceleration with Denoising Diffusion GANs.} Prior work ~\cite{xiao2021tackling} attributes the thousands of denoising steps in diffusion models to the fact that they commonly approximate the denoising distribution with Gaussian noise, thus requiring small step size.  Inspired by  ~\cite{xiao2021tackling} that adopts GAN to model the denoising distribution which enables larger step size and less steps, DiffGAN-TTS~\cite{liu2022diffgan} applies a pretrained GAN as the acoustic generator for acceleration. To further speed up, DiffGAN-TTS~\cite{liu2022diffgan} also introduces an active shallow diffusion mechanism that conduct denoising conditioned on  the  coarse prediction by pretrained GAN. Experimental results show that DiffGAN-TTS~\cite{liu2022diffgan} can generate high-quality audio with only 1 step.

\subsubsection{Adaptive modeling for multi-speaker setting}
\ 

Prior works~\cite{jeong2021diff,popov2021grad} generate Mel-spectrogram 
with a text-conditioned diffusion model. However, they can be trained only when the transcribed data of the target speaker is provided. By applying iterative latent variable sampling ~\cite{choi2021ilvr} to Grad-TTS~\cite{popov2021grad}, Grad-TTS with ILVR~\cite{levkovitch2022zero} proposes to mix the latent variable  with the reference voice from a target speaker during inference, leading to zero-shot speaker without any training. Also follows Grad-TTS~\cite{popov2021grad}, another work Grad-StyleSpeech~\cite{kang2022any}  encodes the mel-spectrogram of reference speech to a styled vector, which is involved in the training of diffusion models. There is another branch of adaptive modeling applying large-scale untranscribed data  in a multi-stage manner.  Guided-TTS~\cite{kim2021guided} proposes a two-stage method that first trains an unconditional DDPM model with  the large-scale untranscribed data, and then generates the mel-spectrogram with the guidance of phoneme classifiers and  the speaker embedding. By contrast,  Guided-TTS 2~\cite{kim2022guided} applies a speaker-conditional DDPM model instead of  the unconditional DDPM model in Guided-TTS~\cite{kim2021guided}. Moreover, Guided-TTS 2~\cite{kim2022guided} adapts the pretrained diffusion model to target speakers with classifier-free guidance and also finetunes the pretrained diffusion model with a short reference speech of the target speaker directly.

\subsubsection{Acoustic models with discrete latent space}
\ 

A branch of prior works~\cite{iashin2021taming,liu2021conditional} proposes to improve the generation efficiency of auto-regressive methods by  compressing the mel-spectrogram into discrete tokens with Vector Quantized Variational Autoencoder (VQ-VAE)~\cite{van2017neural}. Diffsound~\cite{yang2022diffsound} follows this discrete setting but proposes a diffusion-based decoder to generate the tokens in a non-autoregressive manner. To solve the data efficiency problem, Diffsound~\cite{yang2022diffsound} also builds a new text-audio dataset for pertaining, based on the large open-source dataset Audioset~\cite{gemmeke2017audio}. NoreSpeech~\cite{yang2022norespeech} also adopts VQ-VAE~\cite{van2017neural}  in the generation of mel-spectrogram. Different from Diffsound~\cite{yang2022diffsound} working on the discretion of mel-spectrogram, NoreSpeech~\cite{yang2022norespeech} aims to robustly transfer the speaking style to the speech output even when the reference audio contains noise, and thus focuses on the discretion of style features. Specifically, NoreSpeech~\cite{yang2022norespeech} applies  diffusion model to generate \textit{continuous} fine-grained style features conditioned on the noisy reference audio, and then discretes these features with  VQ-VAE~\cite{van2017neural}. The ground truth style features for diffusion model during training are generated by a pretrained style teacher (e.g., GenerSpeech~\cite{huang2022generspeech} and NANSY~\cite{choi2021neural}) on clean audio. With the fine-grained style features, as well as global speaker embedding, NoreSpeech~\cite{yang2022norespeech} generates Mel-spectrogram with the same decoder in ~\cite{huang2022generspeech}. 

\subsubsection{Fine-grained control over audio generation}
\ 

\textbf{Controllable emotional models.} Most attempts require additional optimization to calculate the emotion intensity values, e.g., relative attributes rank (RAR)~\cite{parikh2011relative}. 
However, these methods may cause degeneration of audio quality. By contrast,
EmoDiff~\cite{guo2022emodiff} proposes a soft-labeled guidance technique 
to control the emotion intensity directly. Specifically, EmoDiff~\cite{guo2022emodiff} first trains an unconditional acoustic model, and then trains an emotional classifier on the diffusion trajectory. During inference, the audio is generated under the guidance of the emotional classifier with soft labels. Experimental results show that EmoDiff~\cite{guo2022emodiff} achieves high generation quality while controlling the emotion precisely. 

\subsection{Vocoder}
Neural vocoders generate waveform based on acoustic feature, e.g., Mel-spectrogram. In earlier researches on vocoders until 2020, autoregressive models have been popular in audio generation for their high-quality output samples but suffer from low inference speed. Although non-autoregressive  methods improve the inference speed significantly by reducing sequential steps, there is still an audio quality gap between non-autoregressive  and  autoregressive methods. A summary of the recent works applying diffusion models to Vocoder is shown in Table~\ref{tab:vocoder_results}.

\begin{table*}[!htbp] \centering 
\caption{Experimental results of vocoder based on diffusion model}
\label{tab:vocoder_results}
\resizebox{0.7\textwidth}{!}{
\begin{tabular}{l|c|l|c|l|c|lcccc}
\hline
Methods      & Stage          &    Dataset              &  MOS (↑)& RTF (↓) &PESQ (↑) & STOI (↑)  \\
\hline
WaveGrad~\cite{chen2020wavegrad}  & Vocoder & LJSpeech& 4.47& - & -&-  \\
DiffWave~\cite{kong2020diffwave}  & Vocoder & LJSpeech & 4.44 & - &- &-\\
DDGM~\cite{nachmani2021denoising}  & Vocoder & LJSpeech&- & - &3.308  & 0.969\\
It{\^o}Wave~\cite{wu2022itowave}  & Vocoder & LJSpeech&4.35&- &- &-\\
InferGrad~\cite{chen2022infergrad}    & Vocoder        &   LJSpeech    &  3.97 & - & 3.578 & 0.976
\\
BDDM~\cite{lam2022bddm}  & Vocoder &LJSpeech &  4.48& 0.438 & 3.98 & 0.987\\
\hline
\end{tabular}}
\end{table*}

\subsubsection{Pioneering works}
\ 

WaveGrad~\cite{chen2020wavegrad}  is a pioneering work combing score matching and diffusion models by estimating the gradient of the data log-density, which bridges the audio quality gap between non-autoregressive  and  autoregressive methods. Specifically, WaveGrad~\cite{chen2020wavegrad} proposes  two models variants conditioned on discrete refinement step index and continuous noise level, and find that the continuous variant is more effective and flexible considering the various refinement steps during inference.  WaveGrad~\cite{chen2020wavegrad} can generate high quality samples with only six refinement steps. Another work DiffWave~\cite{kong2020diffwave} is the first model showing a high versatility of waveform generation applications based on diffusion models. In the vocoder task, DiffWave~\cite{kong2020diffwave} is conditioned on mel-spectrogram, and achieves comparable speech quality to the strong autoregressive methods. DiffWave~\cite{kong2020diffwave} can also produce realistic voices and consistent word-level pronunciation in unconditional and class-conditional settings.  

\subsubsection{Towards  efficient vocoders}
\ 

With a shared noise schedule for training and sampling, DDPM~\cite{ho2020denoising} requires thousands of sampling iterations for high-quality generation~\cite{lam2022bddm}. This property inspires investigations to speed up DDPM~\cite{ho2020denoising} by  improving the noise schedule. A branch of work on vocoders applies different noise schedules for training and sampling,  including WaveGrad~\cite{chen2020wavegrad} and DiffWave~\cite{kong2020diffwave}. However, these noise schedules are specifically designed and cannot  easily generalize to other settings~\cite{lam2022bddm}. Another branch of methods searches for a  subsequence of the training schedule along the time axis (e.g., DDIM~\cite{song2020denoising}), while it is still challenging to find a short and effective schedule~\cite{lam2022bddm}.

\textbf{Schedule prediction by additional networks.} In order to find a shorter noise schedule for sampling, ~\cite{lam2022bddm} proposes an additional schedule network to predict the schedule directly. Together with the original score network in DDPM~\cite{ho2020denoising}, the model is termed  bilateral denoising diffusion model (BDDM)~\cite{lam2022bddm}. With both networks, BDDM~\cite{lam2022bddm} converges faster during training with the negligible improvement of computation cost  than  DDPM~\cite{ho2020denoising}. In the vocoder task, BDDM~\cite{lam2022bddm} can generates indistinguishable samples from human speech with only seven steps, 143x and 28.6x faster than WaveGrad~\cite{chen2020wavegrad} and DiffWave~\cite{kong2020diffwave}, respectively.

\textbf{Efficient inference by joint training.} To reduce the inference iterations while maintaining the generation quality, 
InferGrad~\cite{chen2022infergrad} proposes to incorporate the inference process into training with an additional loss. Specifically, InferGrad~\cite{chen2022infergrad} is optimized to minimize the gap between ground-truth samples and samples generated under inference schedules with a few iterations. Evaluated on LJSpeech dataset, InferGrad~\cite{chen2022infergrad} achieves comparable voice quality as WaveGrad~\cite{chen2020wavegrad} with 3x speedup.

\subsubsection{Improvement from statistical perspective}
\ 

\textbf{Improvement with noise prior.} In DDPM~\cite{ho2020denoising}, Gaussian noise is used 
 in the diffusion process which enables sample arbitrary states  without calculating the previous steps. After pointing out that this valuable property stems  from the fact the sum of two Gaussian distributions is still Gaussian distribution, denosing diffusion gamma models(DDGM)~\cite{nachmani2021denoising} claims that Gamma distribution can also meet the adding requirement and benefit DDPM by better fitting the estimated noise than Gaussian noise. Following the noise schedules in WaveGrad~\cite{chen2020wavegrad}, DDGM~\cite{nachmani2021denoising} improves the quality of generated audio than WaveGrad~\cite{chen2020wavegrad}. Quality improvement is also observed in the image generation area.  Another work~\cite{lee2021priorgrad}  points out that the Gaussian noise prior may be insufficient to represent all modes of the samples (e.g., the different voiced and unvoiced segments),  leading to a discrepancy between the real data distribution and the choice of prior and thus the training inefficiency.
Thus, PriorGrad~\cite{lee2021priorgrad} proposes to 
apply an adaptive prior from the data statistics for the efficiency improvement of the conditional diffusion model for speech analysis. Specifically, PriorGrad~\cite{lee2021priorgrad} first computes the mean and variance based on the conditional data, and then maps the computed statistics as the mean and variance of Gaussian prior. With the same mean and variance, the noise is similar to  data distribution at the instance-level. Empirical study shows that PriorGrad~\cite{lee2021priorgrad} can generate high quality outputs with significantly accelerated inference, either waveform for vocoder or Mel-spectrogram for acoustic model.

\textbf{Other improvements.} It{\^o}Wave~\cite{wu2022itowave} is the first to propose a vocoder based on linear It{\^o} SDE. Based on Mel-spectrogram, It{\^o}Wave~\cite{wu2022itowave} 
achieves higher MOS with 95\% confidence than WaveGrad~\cite{chen2020wavegrad} and DiffWave~\cite{kong2020diffwave}. SpecGrad~\cite{koizumi2022specgrad} proposes to adopt the spectral envelope of diffusion noise to the conditional log-mel spectrum,  which improves the sound quality especially for the high-quality bands. 

\subsection{End-to-end frameworks}

Instead of treating acoustic modeling and vocoder modeling as independent processes, a branch of work evolves from partially end-to-end methods to fully end-to-end methods gradually.  Resembling the two-stage frameworks, partially end-to-end methods~\cite{sotelo2017char2wav,ping2018clarinet} also adopt two models as  acoustic model and vocoder, but differentiates by training the two models in a joint manner.  By contrast, (fully) end-to-end frameworks adopt a single model to generate waveform from text without acoustic features as explicit representation. A line of fully end-to-end work adopts an adversarial decoder (or GAN), including FastSpeech 2~\cite{ren2020fastspeech}, EATS~\cite{donahue2020end} and EFTS-Wav ~\cite{miao2021efficienttts}.  Most end-to-end methods still rely on generating mel-spectrogram for text-speech alignment, and a spectrogram-free flow-based method is investigated in Wave-Tacotron~\cite{weiss2021wave} by simply maximizing likelihood. A limitation of Wave-Tacotron~\cite{weiss2021wave} is that the decoder remains autoregressive making it at a disadvantage compared with non-autoregressive counterparts.  In this section, we will discuss the  recent progresses of end-to-end methods based on diffusion model.

\textbf{Pioneering works.} In contrast to WaveGrad~\cite{chen2020wavegrad} converting Mel-spectrogram to waveform,  
WaveGrad 2~\cite{chen2021wavegrad} adopts an end-to-end manner that takes a phoneme sequence as input and generates the audio directly.  Specifically,  WaveGrad~\cite{chen2020wavegrad} decoder is integrated at the end of Tacotron 2 ~\cite{shen2018natural} encoder for a fully differentiable model. As for the duration alignment, WaveGrad 2~\cite{chen2021wavegrad} generates duration information with a non-attentive  Tacotron~\cite{shen2020non} and optimizes a duration predictor towards ground-truth duration during training. By adjusting the refinement steps, WaveGrad 2~\cite{chen2021wavegrad} achieves a trade-off between fidelity and speed. Experimental results show that WaveGrad 2~\cite{chen2021wavegrad} can generate high-quality audio in an end-to-end manner compared to strong baselines. Controllable Raw audio synthesis with High-resolution (CRASH) ~\cite{rouard2021crash} is a concurrent work to WaveGrad 2~\cite{chen2021wavegrad} that proposes an end-to-end model for drum sound  synthesis. Based on SDE, CRASH ~\cite{rouard2021crash} applies a noise-conditioned U-Net to estimate the score function, and introduces a class-mixing sampling to generate 'hybrid' sounds. Experimental results show that CRASH ~\cite{rouard2021crash} can generate drum sounds in multiple tasks, e.g., interpolations and inpainting.

\textbf{Generation of fullband audios.} While previous work focusing on the generation of band-limited audios due to model constraints,   DAG~\cite{pascual2022full} adopts an end-to-end manner to generate full-band audios directly. Based on SDE, DAG~\cite{pascual2022full} introduces an encoder-decoder architecture, which downsamples and upsamples the input sequentially. Experimental results show that DAG~\cite{pascual2022full} improves the   quality and diversity over existing label-conditioned methods.

\textbf{Model based on  It{\^o} SDE.} 
Inspired by  It{\^o}Wave~\cite{wu2022itowave}, It{\^o}n~\cite{shi2022iton}  
proposes an end-to-end model for speech synthesis based on It{\^o} SDE. Apart from the encoder-decoder architecture, It{\^o}n~\cite{shi2022iton} introduces a dual-denoiser structure  for the generation of mel-spectrogram and waveform, respectively. Moreover, It{\^o}n~\cite{shi2022iton} adopts a two-stage training strategy that trains the encoder and Mel denoiser in the first stage, and the wave denoiser in the second stage.

\section{Speech Enhancement}
\label{sec:enhancement}

\begin{table*}[!htbp] \centering 
\caption{Speech enhancement}
\label{tab:stage_papers}
\resizebox{0.9\textwidth}{!}{
\begin{tabular}{l|l|l|l}
\hline
Task & Category & Methods           & Resources \\
\hline
Enhancement by removing &In time-frequency domain & SGMSE~\cite{welker2022speech} & \href{https://github.com/sp-uhh/sgmse}{Code}  \\
&& SGMSE+~\cite{richter2022speech} & \href{https://github.com/sp-uhh/sgmse}{Code} / \href{https://www.inf.uni-hamburg.de/en/inst/ab/sp/publications/sgmse}{Project}   \\
&& Unfolded CD~\cite{yen2022cold}  &  \\
\cline{2-4}
&  In time domain & DiffuSE~\cite{lu2021study}  &        \\
&& CDiffuSE~\cite{lu2022conditional}&\href{https://github.com/neillu23/cdiffuse}{Code}    \\

\cline{2-4}
& Unsupervised restoration  &UVD~\cite{saito2022unsupervised}  &   \\
&& Refiner~\cite{sawata2022versatile} &  \\

\hline
Enhancement by adding& Pioneering work & NU-Wave~\cite{lee2021nu}  & \href{https://github.com/mindslab-ai/nuwave}{Code} / \href{https://mindslab-ai.github.io/nuwave/}{Project}  \\
&& NU-Wave 2~\cite{han2022nu}    & \href{https://mindslab-ai.github.io/nuwave2/}{Project}   \\
\cline{2-4}
& Further improvement&Improved sampling~\cite{yu2022conditioning}                  & \href{https://github.com/yoyololicon/diffwave-sr}{Code} / \href{https://yoyololicon.github.io/diffwave-sr/}{Project}    \\

&& Improved DiffWave~\cite{zhang2021restoring} &  \\
\hline
More various tasks and frameworks & Source separation & DiffSep ~\cite{scheibler2022diffusion}  & \\
\cline{2-4}
& Voice conversion & DiffSVC~\cite{liu2021diffsvc} & \\
\cline{2-4}
 &  Unified framework for multiple tasks  & CQT-Diff~\cite{moliner2022solving} &    \\
& & UNIVERSE~\cite{serra2022universal}  &   \\

\hline 
\end{tabular}}
\end{table*}

Apart from text-to-speech generation, diffusion models have also been widely used in improving the quality of existing degraded audio. 
Numerous factors can cause the degradation of audio quality and we divide them into two classes according to the restoration type. The first branch of methods removes perturbations in the original clean audio, e.g., noise and reverb. The second branch restores missing parts or adds the desired part, e.g., audio super-resolution. Numerous deep speech enhancement methods have been investigated and they can be categorized into two classes. A discriminative method minimizes the difference between enhanced and clean speech~\cite{fu2018end,koizumi2018dnn,fu2019metricgan,defossez2020real}, while generative models are optimized by estimating the
distribution of clean signals~\cite{pascual2017segan,qian2017speech,soni2018time,strauss2021flow,leglaive2020recurrent}. Despite a superior result regarding objective metrics, the discriminative class often suffers from sounding unnatural compared with the generative class. Diffusion model falling into the generative class is a promising method for bridging its gap with the discriminative class.

\subsection{Enhancement by removing}

\subsubsection{Audio restoration in the time-frequency domain}
\

\textbf{A \textit{pure generative} work.}  Although formulated on diffusion models, CDiffuSE~\cite{lu2022conditional} is trained to estimate the difference between clean and noisy speech. Therefore, it is pointed out in~\cite{welker2022speech} that CDiffuSE~\cite{lu2022conditional}  can be considered a discriminative task. To make the method \textit{pure generative} and also avoid any noise prior, SGMSE~\cite{welker2022speech} proposes a  method based on stochastic differential equations (SDE)~\cite{song2019generative, song2020score}. The score function is used to evaluate the quality of the enhanced signal and guide the optimization of the model. In contrast to DiffuSE~\cite{lu2021study} and CDiffuSE~\cite{lu2022conditional} performing diffusion on the time-domain waveform, SGMSE~\cite{welker2022speech} investigates speech enhancement on the time-frequency short-time Fourier transform (STFT) domain to exploit its rich structures. Beyond amplitude of the complex coefficients, it also works on directly enhancing the phase so that an inverse STFT can be applied without extra need of phase retrieval. Experimental results show that SGMSE~\cite{welker2022speech}   maintains more natural structures with fewer artifacts than prior work, and achieves quantitative improvements, e.g., an SI-SAR improvement of 3dB over CDiffuSE~\cite{lu2022conditional}.

\textbf{Further improvement of SGMSE~\cite{welker2022speech}.}  SGMSE+~\cite{richter2022speech} further extends SGMSE~\cite{welker2022speech} with detailed theoretical analysis and improved model architecture. Specifically, SGMSE+~\cite{richter2022speech} presents a theoretical review of the underlying score matching objective, and investigates different sampling configurations during inference. Inspired by the image generation, SGMSE+~\cite{richter2022speech} applies a Noise Conditional Score Network (NCSN++) architecture~\cite{song2020score},  and adapts to the speech area, achieving significant performance improvement over SGMSE~\cite{welker2022speech} on multiple tasks (e.g., speech enhancement and dereverberation). Compared to discriminative methods, SGMSE+~\cite{richter2022speech} achieves comparable results while showing higher generalization capability in a different corpus from that of training. SGMSE+~\cite{richter2022speech} also achieves  state-of-the-art performance in the single-channel dereverberation task.

\textbf{Various forms of degradations.} Compared to the default Gaussian noise in DDPM~\cite{ho2020denoising}, cold diffusion~\cite{bansal2022cold} improves the sampling procedure in the computer vision area and shows impressive generalization capability in a broad family of degradations, e.g., blur, masking, and downsampling. Unfolded CD~\cite{yen2022cold} follows the procedures in ~\cite{bansal2022cold} and adapts cold diffusion to the speech enhancement task. Specifically, ~\cite{yen2022cold} follows the sampling process of ~\cite{bansal2022cold} while modifying the degradation progress by containing the clean sample with an iterative interpolation. As for the training strategy,  ~\cite{yen2022cold}  proposes unfolded training method for better error correction in the degradation and restoration process, improving  the model performance and stability. Experimental results show that ~\cite{yen2022cold} outperforms other diffusion-based models on the speech enhancement task.

\subsubsection{Audio restoration in the time domain}
\ 

DiffuSE~\cite{lu2021study} is a pioneering work to apply diffusion model to speech enhancement, also known as audio denoising. In terms of the basic architecture, DiffuSE is inspired by a diffusion-based waveform generative model DiffWave~\cite{kong2020diffwave}. With the same goal of generating a clean waveform audio, DiffWave~\cite{kong2020diffwave} conditions on a clean mel-spectrogram, while DiffuSE~\cite{lu2021study} relies on noisy audio. A major difference of DiffuSE~\cite{lu2021study} from DiffWave~\cite{kong2020diffwave} lies in its proposed \textit{supportive reverse process}. Specifically, without access to clean spectral features, DiffuSE~\cite{lu2021study} resorts to conditioning on noisy spectral features but pretraining the model with clean spectral features as the conditioner. The supportive reverse process uses a noisy speech Mel-spectral sample instead of isotropic gaussian noise as an initial point for a more efficient clean speech recovery. DiffuSE~\cite{lu2021study} furthers adopts the fast sampling algorithm developed in DiffWave~\cite{kong2020diffwave} to speed up its supportive reverse process.  On the VoiceBankDEMAND dataset, DiffuSE~\cite{lu2021study} achieves performance comparable to other time-domain generative methods. In the formulation of DiffuSE~\cite{lu2021study}, it is assumed that isotropic Gaussian noise is applied in the diffusion as well as its reverse process. However, this assumption does not always hold because the noise characteristics are often non-Gaussian. DiffuSE~\cite{lu2021study} directly combines the noisy speech signal in the sampling process, causing a noise-type mismatch between diffusion and reverse. To address this problem, CDiffuSE~\cite{lu2022conditional} formulates a generalized conditional diffusion model to incorporate the observed noisy data, which facilitates estimating both Gaussian noise and non-Gaussian noise. As a result, CDiffuSE~\cite{lu2022conditional} achieves superior performance and demonstrates great generalization when discriminative approaches fail.  

\subsubsection{Unsupervised restoration}
\ 

\textbf{Unsupervised image restoration with diffusion models.}  Without requiring the pairs of clean and degraded images, unsupervised image restoration~\cite{santurkar2019image,gu2020image,pan2021exploiting,kawar2022denoising} has achieved impressive results, but most of them~\cite{santurkar2019image,gu2020image,pan2021exploiting} suffer from the heavy computations of iterations. To solve the data problem,  denoising diffusion restoration model (DDRM) ~\cite{kawar2022denoising} is the first general sampling-based restoration method that can efficiently generate high-quality, diverse, yet valid solutions for general content images. Specifically, DDRM~\cite{kawar2022denoising} samples the posterior distribution with a linear degradation operator.

\textbf{Unsupervised dereverberation.} Dereverberation of \textit{natural reverb} has been widely studied in prior work, while removing \textit{artificial reverb} is still challenging since the higher number of variations  make  supervised methods cannot generalize to unseen data pairs in the training set. To avoid requiring a large amount of paired  samples and tackle various types of reverb, ~\cite{saito2022unsupervised} proposes an unsupervised vocal dereveration (UVD) method  based on DDRM ~\cite{kawar2022denoising}. Since DDRM~\cite{kawar2022denoising} assumes a known linear degradation operator, UVD~\cite{saito2022unsupervised}  extends DDRM~\cite{kawar2022denoising}  to music dereverberation task where the reververation operators are always unknown. Specifically, the operator is initially estimated and adaptively corrected by  weighted prediction error (WPE)~\cite{nakatani2010speech}. The predicted clean (dry) signal from a diffusion model is used for operator correction, and the estimated dry signals are generated after iterations. Experimental results show that  UVD~\cite{saito2022unsupervised} outperforms unsupervised and supervised benchmarks in both objective and subjective evaluations. 

\textbf{Two-stage refinement in speech enhancement.}  Research on speech enhancement has achieved significant improvement in terms of signal-to-noise (SNR) ratio but sometimes degrades the speech quality (e.g., naturalness), leading to the degradation of downstream applications. To remove the distortions of speech enhancement outputs, Refiner ~\cite{sawata2022versatile} applies a  diffusion model pretrained on clean speech data  to detect the degraded part,  and then replaces them with newly generated clean ones in the manner of denoising diffusion restoration models(DDRM)~\cite{kawar2022denoising}. Experimental results show that Refiner ~\cite{sawata2022versatile} is versatile since it improves speech quality with regard to various speech enhancement methods. Moreover, the Refiner ~\cite{sawata2022versatile} can also be integrated into the speech enhancement model for joint optimization in the future.

\begin{table}[!htbp] \centering 
\caption{AUDIO RESTORATION RESULTS}
\label{tab:enhancement_results}
\resizebox{0.98\textwidth}{!}{
\begin{tabular}{l|l|l|l|l|l|l|l|l|l|l|l}
\hline
Methods&  Training set& Testing set& POLQA $\uparrow$ & PESQ $\uparrow$& ESTOI $\uparrow$& SI-SDR[dB] $\uparrow$& SI-SIR[dB] $\uparrow$& SI-SAR[dB] $\uparrow$& CSIG $\uparrow$& CBAK $\uparrow$& COVL $\uparrow$ \\
\hline 
CDiffuSE~\cite{lu2022conditional}& WSJ0-CHiME3& WSJ0-CHiME3 & 2.77$\pm$0.52& 2.15$\pm$0.49& 0.80$\pm$0.09& 7.3$\pm$1.9& 19.5$\pm$1.6& 7.8$\pm$1.8& $-$& $-$& $-$  \\
SGMSE~\cite{welker2022speech}& WSJ0-CHiME3& WSJ0-CHiME3 &2.98$\pm$0.60& 2.28$\pm$0.57& 0.86$\pm$0.09& 14.8$\pm$4.3& 25.4$\pm$5.6& 15.3$\pm$4.2& $-$& $-$& $-$ \\
SGMSE+~\cite{richter2022speech}& WSJ0-CHiME3& WSJ0-CHiME3 & 3.73$\pm$0.53& 2.96$\pm$0.55& 0.92$\pm$0.06& 18.3$\pm$4.4& 31.1$\pm$4.6& 18.6$\pm$4.5& $-$& $-$& $-$ \\

\hline
CDiffuSE~\cite{lu2022conditional}& VBD& WSJ0-CHiME3 & 2.20$\pm$0.50& 1.84$\pm$0.41& 0.71$\pm$0.10& 3.8$\pm$2.5& 21.6$\pm$7.0& 4.0$\pm$2.5& $-$& $-$& $-$ \\
SGMSE~\cite{welker2022speech}& VBD& WSJ0-CHiME3 &2.31$\pm$0.44& 1.62$\pm$0.31& 0.74$\pm$0.11& 10.5$\pm$4.1& 20.2$\pm$7.6& 11.4$\pm$3.5& $-$& $-$& $-$ \\
SGMSE+~\cite{richter2022speech}& VBD& WSJ0-CHiME3 & 3.43$\pm$0.61& 2.48$\pm$0.58& 0.90$\pm$0.07& 16.2$\pm$4.1& 28.9$\pm$4.6& 16.4$\pm$4.1& $-$& $-$& $-$ \\

\hline
DiffuSE~\cite{lu2021study}& VBD& VBD& $-$ &2.43 & $-$ & 10.5$\pm$0.14& 30.0$\pm$0.71& 10.8$\pm$0.11& 3.63& 2.81& 3.01 \\

CDiffuSE~\cite{lu2022conditional}& VBD&  VBD & $-$& 2.52& 0.79& 12.6/12.1$\pm$0.10& 28.2$\pm$0.36& 12.3$\pm$0.09& 3.72& 2.91& 3.10  \\

SGMSE~\cite{welker2022speech}& VBD& VBD& $-$& 2.28& 0.80& 16.2/15.1$\pm$0.27& 24.9$\pm$0.42& 15.7$\pm$0.25& $-$& $-$ & $-$  \\

SGMSE+~\cite{richter2022speech}& VBD& VBD& $-$& 2.93& 0.87& 17.3& $-$& $-$& $-$& $-$ & $-$  \\

UVD~\cite{saito2022unsupervised}& NHSS& NHSS& $-$& $-$& $-$& $-$& $-$& $-$& $-$& $-$& $-$ \\ 

Unfolded CD~\cite{yen2022cold}& VBD & VBD& $-$& 2.77& $-$& $-$& $-$& $-$& 3.91& 3.32& 3.33 \\

Refiner~\cite{sawata2022versatile}& VBC& VBC& $-$& $-$& $-$& $-$& $-$& $-$& $-$& $-$& $-$\\
\hline
\end{tabular}
}
\end{table}

\subsection{Enhancement by adding}

\subsubsection{Pioneering works on  audio super-resolution}
\ 

Audio super-resolution~\cite{kuleshov2017audio,lim2018time}, also widely known as upsampling~\cite{pons2021upsampling} or bandwidth extension~\cite{kuleshov2017audio}, aims to generate audio of a high sampling rate from that of a low sampling rate via extending its bandwidth. Compared to previous work with 16kHz as the target frequency~\cite{li2015dnn,kuleshov2017audio,lim2018time,li2019speech,kim2019bandwidth,birnbaum2019temporal,hou2020speaker,pons2021upsampling}, NU-Wave~\cite{lee2021nu} is the first to synthesize 48kHz waveforms from 16kHz or 24kHz inputs, and also the first to apply diffusion models for audio super-resolution. Following  the  structures of  prior diffusion-based vocoders (DiffWave~\cite{kong2020diffwave} and WaveGrad~\cite{chen2020wavegrad}), NU-Wave~\cite{lee2021nu} adapts the model to  audio super-resolution task, e.g, introducing a sinusoidal 128-dimensional vector as noise level embedding. Moreover, NU-Wave~\cite{lee2021nu} empirically finds that the receptive field for a conditional signal should be larger than that of  the noisy input, and thus modifies the local conditioner for the information ensemble of signals with different receptive fields and  upscaling ratios. Experimental results show that NU-Wave~\cite{lee2021nu} outperforms the baseline methods in all cases, even with smaller model capacity.

NU-Wave 2~\cite{han2022nu} further improves NU-Wave~\cite{lee2021nu} from two aspects. On the one hand, NU-Wave 2~\cite{han2022nu} adopts short-time Fourier convolution (STFC) to overcome the limitations of NU-Wave~\cite{lee2021nu} that fails to generate   harmonics of vowels ~\cite{lee2021nu} and various frequency bands~\cite{liu2021voicefixer, nguyen2022tunet}. On the other hand, different from prior models that the initial and target sampling rates are fixed,
NU-Wave 2~\cite{han2022nu} defines a new task \textit{general neural audio upsampling} that the inputs can be any sampling rate for a single model. Specifically, NU-Wave 2~\cite{han2022nu}  proposes bandwidth spectral feature transform (BSFT) layer for  adapting to  various sampling rates. Experimental results show that NU-Wave 2~\cite{han2022nu} 
has many valuable characteristics compared to baseline methods,  e.g., only NU-Wave 2~\cite{han2022nu} can generate harmonics.

\textbf{Improved sampling method.} Diffusion-based audio super-resolution is commonly conducted by conditioning the denoising network on low-resolution audio. ~\cite{yu2022conditioning} further improves the audio quality by injecting the low-resolution audio into the sampling process as a condition if the downsampling schedule is known. This sampling method can be used directly in other diffusion-based super-resolution methods. Moreover, ~\cite{yu2022conditioning} can generalize to multiple settings with a UDM, e.g., varying upscaling ratios. Experimental results show that ~\cite{yu2022conditioning}  achieves the state-of-the-art log-spectral-distance (LSD) on the 48kHz VCTK Multi-Speaker benchmark.

\textbf{Improved model architecture.} ~\cite{zhang2021restoring} improves the speech quality caused by  deterministic mathematical degradation, e.g., compression, clipping and downsampling. Conditioned on degraded Mel-spectrogram,  ~\cite{zhang2021restoring} 
finds that DiffWave~\cite{kong2020diffwave} can restore the degraded speech to some extent. For further improvement, ~\cite{zhang2021restoring} replaces the original upsampler  of DiffWave~\cite{kong2020diffwave} with a CNN sampler and trains 
 the CNN sampler  seperately with the original upsampler as the reference. We summarize the experimental results of speech enhancement models in Table~\ref{tab:enhancement_results}.

\subsection{Miscallenous audio tasks}
\textbf{Source separation.} Source separation is a task that aims  to recover speech of interest from a mixed signal~\cite{scheibler2022diffusion}. 

Early work on source separation is based on short-time Fourier transform (STFT), which aims to get the clean spectrum from the mixed spectrum. However, using STFT in audio separation is not sensitive to phase and needs high-frequency resolution, making it low accuracy of reconstruction and time-consuming. To overcome these problems, other time-domain approaches are proposed, including independent component analysis (ICA)~\cite{choi2005blind}, non-negative matrix factorization(NMF)~\cite{yoshii2013beyond}.DiffSep~\cite{scheibler2022diffusion} is the first diffusion-based method for single-channel speech separation. Specifically, DiffSep~\cite{scheibler2022diffusion} trains a SDE model that converts the separated signal to the mixtures, thus the reverse process can separate individual source from the mixed signal.  DiffSep ~\cite{scheibler2022diffusion}  can also be applied for audio denoising since noise can be seen as a type of extra source. 

\textbf{Voice conversion.} Voice Conversion edits the source speakers to adapt to the target speaker by changing speech signal features. Traditional techniques include vector quantization~\cite{ shikano1986speaker,abe1990voice}, hidden Markov models~\cite{tokuda1995speech, kim1997hidden}, and Gausian mixture models~\cite{stylianou1998continuous, veaux2011intonation}. However, these methods only focus on specific parts of spectrum rather than the entire spectrum, resulting in poor speech quality. To solve these problems,~\cite{lai2016phone, sun2015voice, pascual2016multi, mobin2016voice, liu2021fastsvc } adopt neural networks which cover the entire spectrum, and DiffSVC~\cite{liu2021diffsvc} is a pioneering work that leveraging diffusion models on voice conversion. Specifically, DiffSVC~\cite{liu2021diffsvc} is developed for singing voice conversion (SVC), which first transforms the phonetic posteriorgrams (PPGs) to spectral features, and then transforms them into waveforms by a trained neural vocoder. This method can get the target speech by any other speech as input (i.e., any-to-one SVC). 

\textbf{Towards a Unified framework for multiple tasks.} A branch of work aims to solve multiple tasks with a single model~\cite{moliner2022solving,serra2022universal}. CQT-Diff ~\cite{moliner2022solving} explores whether a pretrained model can be used for different tasks without knowing the degradation type during inference. Specifically, CQT-Diff ~\cite{moliner2022solving} excels at the bandwidth extension task, and achieves competitive results to baseline methods in long audio inpainting and declipping tasks without retraining. Note that the method is termed as   CQT-Diff since Constant-Q Transform (CQT) is adopted for  strong harmonic signals (e.g., music). Except common background noise and reverberation, another work ~\cite{serra2022universal} points out that a large number of distortions may exist in the audios, including bandwidth reduction, clipping, silent gaps.Therefore, ~\cite{serra2022universal} proposes a universal system that can tackle 55 different distortions in an end-to-end manner. With a conditioner network that tackles the main part of a task, a generator network is formulated on the score-based diffusion models for waveform synthesis. Experimental results show that this approach achieves the state-of-the-art performance according to a subjective text with expert listeners~\cite{serra2022universal}.

\section{Conclusion}
\label{sec:conclusion}

This survey reviews the recent progress of speech synthesis based on diffusion model. After introducing the background of audio signals and diffusion model, we discuss recent works of two main tasks in the speech synthesis field: text-to-speech and  speech enhancement. We hope this survey could provide an insightful understanding for the researchers who are interested in speech synthesis as well as generative AI.

\bibliographystyle{ACM-Reference-Format}
\bibliography{bib_mixed,bib_local}

\end{document}